# Inverse orbital Hall effect and orbitronic terahertz emission observed in the materials with weak spin-orbit coupling

Ping Wang,[1] Zheng Feng,[2] Yuhe Yang,[1] Delin Zhang,[1,*] Quancheng Liu,[3] Zedong Xu,[1] Zhiyan Jia,[1] Yong Wu,[4] Guoqiang Yu,[5] Xiaoguang Xu,[4] Yong Jiang[1,*]

[1] Institute of Quantum Materials and Devices, School of Electronic and Information Engineering; State Key Laboratory of Separation Membrane and Membrane Processes, Tiangong University, Tianjin, 300387, China.

[2] Microsystem & Terahertz Research Center, CAEP, Chengdu, 610200, China.

[3] School of Information Engineering, Southwest University of Science and Technology, Mianyang, 621010, China.

[4] School of Materials Science and Engineering, University of Science and Technology Beijing, Beijing, 100083, China.

[5] Beijing National Laboratory for Condensed Matter Physics, Institute of Physics, Chinese Academy of Sciences, Beijing, 100190, China.

These authors contributed equally: Ping Wang, Zheng Feng, Yuhe Yang, Delin Zhang.

*Corresponding authors. Email: zhangdelin@tiangong.edu.cn (D.L.Z.), and yjiang@tiangong.edu.cn (Y.J.)

Abstract:

The Orbital Hall effect, which originates from materials with weak spin-orbit coupling, has attracted considerable interest for spin-orbitronic applications. Here, we demonstrate the inverse effect of the orbital Hall effect and observe orbitronic terahertz emission in the Ti and Mn materials. Through spin-orbit transition in the ferromagnetic layer, the generated orbital current can be converted to charge current in the Ti and Mn layers via the inverse orbital Hall effect. Furthermore, the inserted W layer provides an additional conversion of the orbital-charge current in the Ti and Mn layers, significantly enhancing the orbitronic terahertz emission. Moreover, the orbitronic terahertz emission can be manipulated by cooperating with the inverse orbital Hall effect and the inverse spin Hall effect in the different sample configurations. Our results not only discover the physical mechanism of condensed matter physics but also pave the way for designing promising spin-orbitronic devices and terahertz emitters.

## INTRODUCTION

The materials with spin-orbit coupling (SOC) possess two distinct channels of angular momentum, namely spin angular momentum (**S**) and orbital angular momentum (**L**), which generate spin current and orbit current in a transverse direction with an applied electric field[1-8]. Recently, charge-spin and spin-charge conversions have been demonstrated in the heavy metals (HM) (e.g. Ta, W) and quantum topological materials (e.g. topological insulators and semimetals) with strong SOC through spin Hall effect (SHE) or Rashba-Edelstein effect and inverse spin Hall effect (ISHE), respectively[1-4,9-12]. These effects have opened up potential technical applications, including spin-orbit torque (SOT) devices and spintronic terahertz (THz) emitters[11-18]. However, due to the domination of the spin contribution in the nonmagnetic materials (NM) with strong SOC, the orbital contribution has not been given much attention. Recently, the orbital Hall effect (OHE) has been theoretically and experimentally observed in NM with weak SOC, where the efficient orbital current is obtained with an applied electric field[7,8,19]. Through OHE, the charge current ($J_C$) can be converted to orbital current ($J_L$), and then to spin current ($J_S$), which can exert an orbital torque to switch the magnetization of the ferromagnetic layer, thereby operating the orbitronic devices[20-24]. However, the partner of OHE, inverse orbital Hall effect (IOHE), as well as the orbitronic THz emitters exhibiting advantages of high emission efficiency, ultra-broad bandwidth, excellent performance, and flexible tunability remains elusive[9,16-18,25,26]. To remedy this, we choose Ti and Mn as the OHE source to study IOHE and the efficiency of orbitronic THz emission originating from IOHE.

## RESULTS AND DISCUSSION

**Inverse orbital Hall effect**

The SHE and ISHE phenomena involve the conversion of $J_C{\rightarrow}J_S$ and $J_S{\rightarrow}J_C$ in heavy metals or quantum materials with strong SOC[3,10], where a transverse flow of spin angular momentum and voltage are generated, respectively, as shown in Fig. 1a. The direction of $J_S$, $J_C$ and the conversion efficiency of $J_C{\rightarrow}J_S$, $J_S{\rightarrow}J_C$ depend on the sign and value of spin Hall angle ($\theta_{SH}$), as given by the equations $J_S \sim \theta_{SH}\ \sigma{\times}J_C$ and $J_C \sim \theta_{SH}\ J_S{\times}\sigma$, where $\sigma$ is the spin polarization[2,4,16]. However, the OHE and IOHE phenomena refer to the conversion of $J_C{\rightarrow}J_L$ and $J_L{\rightarrow}J_C$, respectively, in NM with weak SOC, resulting in a transverse flow of orbital angular momentum and voltage, as depicted in Fig. 1b. Through OHE, $J_C$ is converted to $J_L$ in the NM layer without relying on the strong SOC. $J_L$ is then transferred to $J_S$, which generates orbital torque in the FM layer due to its SOC, as widely reported[20-24]. IOHE is the inverse process of OHE, where $J_S$ generated in the FM layer first converts to $J_L$, then flows into the NM layer and converts to $J_C$. The direction of $J_L$, $J_C$ and the conversion efficiency of $J_C{\rightarrow}J_L$, $J_L{\rightarrow}J_C$ depend on the sign and value of orbital Hall angle ($\theta_{OH}$), which is given by the equations $J_L \sim \theta_{OH}\ \sigma_{OH}{\times}J_C$ and $J_C \sim \theta_{OH}\ J_L{\times}\sigma_{OH}$, where $\sigma_{OH}$ is the orbital polarization in OHE. $\theta_{OH} = \rho\ \sigma_L\ (2e/\hbar)$ is similar to $\theta_{SH} = \rho\ \sigma_S\ (2e/\hbar)$, where $\sigma_S$, $\sigma_L$, and $\rho$ are spin and orbital Hall conductivity and resistivity, respectively[8].

To investigate IOHE as proposed above, we designed and fabricated the bilayered structures of Co (2 nm)/$X$ (4-60 nm) ($X$ = Ti, Mn) and carried out the experiments using THz emission spectroscopy[18]. The Co layer used here can effectively achieve $J_S{\rightarrow}J_L$ conversion due to its relatively large spin-orbital conversion efficiency ($C_{Co}$), where the sign and value of $C_{Co}$ depend on the spin-orbit correlation $R = \langle\mathbf{L}\ \mathbf{S}\rangle$ of the Co layer[20,21,27]. Meanwhile, Ti and Mn possess large OHE as demonstrated in theory and experiment[7,8,19,21,28,29]. Figure 2a illustrates the physical mechanism of the $J_S{\rightarrow}J_L{\rightarrow}J_C$ conversion in the Co/$X$ ($X$ = Ti, Mn) bilayered structures. When a femtosecond (*fs*) laser is pumped into the Co layer, both $\mathbf{J_S}$ and $\mathbf{J_L}$ are simultaneously generated due to SOC of the Co layer (The possible contribution of $J_L$ can be directly generated in the Co layer by the laser). $\mathbf{J_L}$ will flow into the Ti or Mn layers and convert to $\mathbf{J_{C,IOHE}}$ due to IOHE (indicated by the blue arrow), while $\mathbf{J_S}$ will directly convert to $\mathbf{J_{C,ISHE}}$ due to ISHE in the Ti or Mn layers (labeled by the red arrow). The phase of $\mathbf{J_{C,ISHE}}$ and the conversion efficiency of $\mathbf{J_S} \rightarrow \mathbf{J_{C,ISHE}}$ depend on the sign and value of $\theta_{SH}$ of the Ti or Mn layers ($\theta_{SH,Ti/Mn} < 0$)[8,30], which can be identified from the polarity and amplitude of the THz signal[16]. Compared to ISHE, the phase and amplitude of $\mathbf{J_{C,IOHE}}$ depend on the sign of $C_{Co}\theta_{OH,Ti/Mn}$, as it arises from the conversion of $\mathbf{J_S} \rightarrow \mathbf{J_L} \rightarrow \mathbf{J_{C,IOHE}}$ (where $C_{Co} > 0$ and $\theta_{OH,Ti/Mn} > 0$)[8,19,20,31]. In the bilayered structures of Co/*X* (where *X* = Ti, Mn), the phase of $\mathbf{J_{C,ISHE}}$ and $\mathbf{J_{C,IOHE}}$ is opposite, and this can be expressed in the polarity of the THz signal. Although $\mathbf{J_{C,ISHE}}$ arises from *X* (where *X* = Ti, Mn), it is negligible owing to the very small $\theta_{SH,Ti/Mn}$ of *X* (*X* = Ti, Mn)[30]. Therefore, $\mathbf{J_{C,IOHE}}$ originating from Ti or Mn dominates the THz signal. Figure 2b and Supplementary Figure 1 show the normalized THz signals of the Co (2 nm)/Ti (4-60 nm) and Co (2 nm)/Mn (4-60 nm) structures with different thicknesses (*d*) of Ti and Mn (the normalized THz signals are obtained by normalizing the original THz signals to the laser absorbance of FM layer and the THz radiation impendence, which represents the spin/orbit-charge conversion efficiency[9,32], see details in Methods). The obvious THz signals can be observed in the Co (2 nm)/Ti (4-60 nm) and Co (2 nm)/Mn (4-20 nm) structures, suggesting that the $\mathbf{J_S} \rightarrow \mathbf{J_C}$ conversion occurs. The polarity of the THz signals is opposite when the samples were flipped due to the opposite direction of $\mathbf{J_L}$ during the measurement (see Supplementary Fig. 2), further confirming the $\mathbf{J_L} \rightarrow \mathbf{J_C}$ conversion in the Co/Ti and Co/Mn structures. To eliminate the contribution of the substrate, Co, Mn, Ti, or MgO layers, we performed THz experiments on these layers, and the results are presented in

Supplementary Fig. 3. We found that the contribution of the THz signals from these layers is negligible. Moreover, to confirm the $\mathbf{J_S} \rightarrow \mathbf{J_C}$ conversion in the Co/Ti and Co/Mn structures due to IOHE, we measured the THz signals of the Co (2 nm)/W (2 nm) and Co (2 nm)/Pt (2 nm) bilayered structures, where $\mathbf{J_{C,ISHE}}$ mainly originates from ISHE in the W and Pt layers[16]. If the THz signals of the Co (2 nm)/Ti (4 nm) and Co (2 nm)/Mn (4 nm) bilayered structures come from ISHE, the polarity of the THz signals will be the same as that of the Co (2 nm)/W (2 nm) bilayered structure but opposite to that of the Co (2 nm)/Pt (2 nm) bilayered structure. This is because Ti, Mn and W have the same negative sign of $\theta_{SH}$, while Pt has a positive sign of $\theta_{SH}$ ($\theta_{SH,Ti}$ = -3.6×10$^{-4}$, $\theta_{SH,Mn}$ = -1.9×10$^{-3}$, $\theta_{SH,W}$ = -3.3×10$^{-1}$, $\theta_{SH,Pt}$ = 1.2×10$^{-1}$)[30,33,34]. However, the polarity of the THz signals in the Co/Ti and Co/Mn bilayered structures is opposite to that of the Co/W bilayered structure but the same as that of the Co/Pt bilayered structure, as shown in Fig. 2c. The polarity of the THz signals for the Co/Ti and Co/Mn bilayered structures is consistent with the sign of $C_{Co}\theta_{OH,Ti/Mn}$, indicating that the THz signals in these structures are induced by IOHE. In addition, the absolute values of the normalized THz peak-to-peak signals (THz ΔPK, which represents the difference between the first peak and second peak) of the Co (2 nm)/Ti (4-60 nm) bilayered structures first increase up to the maximum at $d_{Ti}$ = 40 nm and then decrease, persisting even at $d_{Ti}$ = 60 nm, as depicted in Fig. 2d. This behavior is in agreement with the long orbital diffusion length of Ti ($\lambda_{OD,Ti}$)[19], as reported previously. In contrast to Ti, Mn has a relatively short orbital diffusion length ($\lambda_{OD,Mn}$).

**Enhanced orbitronic THz emission assisted by efficient spin-orbital current conversion**

To gain a deeper understanding of the crucial $\mathbf{J_S} \rightarrow \mathbf{J_L} \rightarrow \mathbf{J_C}$ conversion and the orbitronic THz emission based on IOHE, as well as the additional contribution from ISHE, we introduced inserted layers (W, Ti, Mn) with different SOC between the Co and $X$ ($X$= Ti, Mn) layers. The normalized THz signals of the Co/$M$/$X$ ($M$=W, Ti, Mn; $X$= Ti, Mn) structures are presented in Fig. 3 and Supplementary Fig. 4. When 2-nm W with a strong SOC ($R_W < 0$, $C_W < 0$) is inserted[20,33], the amplitude of the THz signals for the Co (2 nm)/W (2 nm)/$X$ (4 nm) ($X$ = Ti, Mn) structures can be significantly enhanced compared to the Co (2 nm)/$X$ (4 nm) ($X$ = Ti, Mn) and Co (2 nm)/W (2 nm) bilayered structures, as shown Fig. 3a. Furthermore, Fig. 3a displays the corresponding difference of absolute value ($|\Delta PK_{(Co/W/Ti)-(Co/W)}|$ and $|\Delta PK_{(Co/W/Mn)-(Co/W)}|$) of the THz $\Delta PK$ between the Co (2 nm)/W (2 nm)/$X$ (4 nm) ($X$ = Ti, Mn) and Co (2 nm)/W (2 nm) structures, as well as the THz $|\Delta PK|$ of the Co (2 nm)/$X$ (4 nm) ($X$ = Ti, Mn) bilayered structures. We have observed that $|\Delta PK_{(Co/W/Ti)-(Co/W)}|$ and $|\Delta PK_{(Co/W/Mn)-(Co/W)}|$ exhibit similar magnitudes, which are more than one order of the magnitude larger than that of the Co (2 nm)/$X$ (4 nm) ($X$ = Ti, Mn) bilayered structures. These results indicate that the inserted W layer not only generates **J**$_{C,ISHE}$ (as indicated by the red arrow in Fig. 3b), but also contributes to **J**$_{C,IOHE}$ (as indicated by the blue arrow in Fig. 3b) in the Co (2 nm)/W (2 nm)/$X$ (4 nm) ($X$= Ti, Mn) structures. The inserted W layer provides an additional large **J**$_{L,W}$ resulting from the coupling between **L** and **S** due to its large SOC (**J**$_L$ from the Co layer is smaller compared to the W layer and is not labeled in the figure). **J**$_{L,W}$ enters the Ti or Mn layers and converts to **J**$_{C,IOHE}$ by IOHE (**J**$_{C,IOHE}$ ~ $C_W$ $\theta_{OH,Ti/Mn}$ **J**$_S$). **J**$_{C,IOHE}$ and **J**$_{C,ISHE}$ in the Co (2 nm)/W (2 nm)/$X$ (4 nm) ($X$ = Ti, Mn) structures have the same phase. Thus, the orbitronic THz emission of the Co (2 nm)/W (2 nm)/$X$ (4 nm) ($X$ = Ti, Mn) structures primarily originates from **J**$_C$ ~ $\theta_{SH,W}$ **J**$_S$ + $C_W$ $\theta_{OH,Ti/Mn}$ **J**$_S$. However, when the Ti (4 nm) and Mn (4 nm) layers with weak SOC ($R_{Ti,Mn} < 0$) are inserted, the $|\Delta PK_{(Co/Mn/Ti)-(Co/Mn)}|$ and $|\Delta PK_{(Co/Ti/Mn)-(Co/Ti)}|$ values for the Co (2 nm)/Ti (4 nm)/Mn (4 nm) and Co (2 nm)/Mn (4 nm)/Ti (4 nm) structures, respectively, are smaller than those of the Co (2 nm)/$X$ (4 nm) ($X$ = Ti, Mn) structures, as depicted in Fig. 3c. In the Co/Ti/Mn and Co/Mn/Ti structures, the conversion of $\mathbf{J_S} \rightarrow \mathbf{J_L}$ predominantly occurs in the Co layer, and $\mathbf{J_L}$ further converts to $\mathbf{J_C}$ in the Ti and Mn layers without an additional $\mathbf{J_L}$ like in the W layer. Therefore, the orbitronic THz emission of the Co/Ti/Mn and Co/Mn/Ti structures still primarily originates from the conversion of $\mathbf{J_S} \rightarrow \mathbf{J_L}$ by the SOC of the Co layer and $\mathbf{J_L} \rightarrow \mathbf{J_C}$ by IOHE of the Ti and Mn layers, as shown in Fig. 3d.

**Collaboration between inverse orbital and spin Hall effects**

Inspired by the physical mechanism of the Co/W/*X* (*X* = Ti, Mn) structures, we investigated the orbitronic THz emission of the different structures: Co (2 nm)/Ti (4-100 nm)/W (2 nm) and Ti (4-100 nm)/Co (2 nm)/W (2 nm), derived from the combined contribution of IOHE and ISHE. The conversion process is illustrated in Figs. 4a and 4c. When the Ti layer is inserted between the Co and W layers, $\mathbf{J_S}$ from the Co layer will be divided into two parts, one part will transfer into the W layer through the Ti layer and convert to $\mathbf{J_{C,ISHE}}$ due to ISHE, the other part will convert to $\mathbf{J_L}$ and flow into the Ti layer. $\mathbf{J_L}$ will then convert to $\mathbf{J_{C,IOHE}}$ due to IOHE. $\mathbf{J_{C,ISHE}}$ and $\mathbf{J_{C,IOHE}}$ have a 180-degree phase shift, as shown in Fig. 4a. During this process, $\mathbf{J_L}$ from the W layer is not induced. Therefore, we did not observe the enhanced amplitude of the THz signals in the Co/Ti/W structure, as we did in the Co/W/*X* (*X* = Ti, Mn) structures. Additionally, we found that the polarity and amplitude of the THz signals can change in the Co/Ti/W structure with an increase in $d_{Ti}$, as presented in Fig. 4b. The reason is that the Co (2 nm)/Ti (10 nm) structure has a larger THz |ΔPK| than that of the Co (2 nm)/Ti (4 nm) structure (see Fig. 2b), and $\mathbf{J_{C,ISHE}}$ and $\mathbf{J_{C,IOHE}}$ have a 180 degree phase shift. Thus the THz |ΔPK| of the Co (2 nm)/Ti (4 nm)/W (2 nm) structure is larger than that of the Co (2 nm)/Ti (10 nm)/W (2 nm) structure. However, when $d_{Ti} > \lambda_{SD,Ti}$, $\mathbf{J_S}$ from the Co layer is blocked by the Ti layer and cannot convert to $\mathbf{J_{C,ISHE}}$ in the W layer. In this case, $\mathbf{J_{C,IOHE}}$ from the Ti layer dominates the THz signals, resulting in small THz |ΔPK|

with the opposite polarity compared to the Co (2 nm)/Ti (4 nm)/W (2 nm) and Co (2 nm)/Ti (10 nm)/W (2 nm) structures in the Co (2 nm)/Ti (40-100 nm)/W (2 nm) structure. In the Ti/Co/W structure, the THz polarity is identical to that of the Co/W structure. The amplitude of the THz emission first increases and then decreases with the increasing $d_{Ti}$, reaching maximum amplitude (much larger than that of the Co/W structure) at $d_{Ti}$ = 10 nm, as shown in Fig. 4d. The THz signals primarily originate from $\mathbf{J_S} \rightarrow \mathbf{J_{C,W}}$ by ISHE of the W layer and $\mathbf{J_S} \rightarrow \mathbf{J_L} \rightarrow \mathbf{J_{C,Ti}}$ by IOHE of the Ti layer. $\mathbf{J_{C,ISHE}}$ from the W layer and $\mathbf{J_{C,IOHE}}$ from the Ti layer have no phase shift as plotted in Fig. 4c. Therefore, the THz |ΔPK| increases first and then decreases with increasing $d_{Ti}$. Additionally, we found that the amplitude of the THz emission also depends on the thickness of the nonmagnetic and ferromagnetic layers[9], besides the cooperation or competition between ISHE and IOHE.

We have demonstrated the occurrence of IOHE in materials with weak spin-orbit coupling such as Ti and Mn, and observed the orbitronic terahertz emission in the Co/Ti and Co/Mn structures. The introduction of a W layer has been found to significantly enhance the orbitronic terahertz emission by inducing an additional conversion of the orbital-charge current. Furthermore, the manipulation of orbitronic THz emission has been achieved by designing specific structural configurations that cooperate with IOHE and ISHE. This study not only helps to explore the physical mechanism of IOHE, but also provides guidance for the design and fabrication of spin-orbitronic devices and THz emitters.

**METHODS**

**Sample preparation**

All the samples in this study were prepared on the $Al_2O_3$ (0001) single crystal substrates using an ultrahigh vacuum magnetron sputtering system at room temperature. To prevent oxidation, a 5-nm MgO capping layer was deposited on the samples. During the deposition process, the working gas Ar was set to 2.5 mTorr. Co (2 nm)/$X$ (4-60 nm) ($X$ = Ti, Mn) structures were used to investigate IOHE. Co (2 nm)/W (2 nm)/$X$ (4-100 nm) ($X$ = Ti, Mn) structures were utilized to study the enhancement of IOHE by inserting materials with strong SOC. Co/Ti (4-100 nm)/W (2 nm) and Ti (4-100 nm)/Co (2 nm)/W (2 nm) structures were employed to understand the collaboration and competition between IOHE and ISHE. The Co (2 nm)/W (2 nm), Co (2 nm)/Pt (2 nm), Co (2 nm)/MgO (5 nm), Ti (4 nm)/MgO (5 nm), Mn (4 nm)/MgO (5 nm), MgO (5 nm) structures were prepared as the reference samples.

**THz emission measurement**

The THz emission measurements were conducted using a home-made THz emission spectroscopy setup that utilized a Ti: sapphire laser oscillator with a center wavelength of 800 nm, a pulse duration of 100 *fs*, an average power of 2 W, and repetition rate of 80 MHz. The *fs* laser beam was split into a pump and a probe beam. The pump beam was directed onto the sample under normal incidence to excite it, and the THz waves generated were detected using the electro-optic sampling technique with the probe beam. For detection, a 2-mm thick ZnTe (110) electro-optic crystal was used. The samples were subjected to an in-plane magnetic field of 50 mT. All measurements were conducted at room temperature in a dry-air environment.

**Normalization of the THz signals**

The normalized THz signals can be obtained by normalizing the original THz signals ($E_{THz}$) to the laser absorbance ($A_{FM}$) of the FM layer and the THz radiation impendence (Z). The laser absorbance ($A_{FM}$) of the FM layer is defined by the equation[32]:

$$A_{FM} = A_{total} \frac{t_{FM}}{t_{total}} \quad (1)$$

where $A_{total}$ and $t_{total}$ are the total laser absorbance and total thickness of the FM/NM bilayer or FM/NM$_1$/NM$_2$(NM$_1$/FM/NM$_2$) trilayer, $t_{FM}$ is the thickness of the FM layer. By measuring the total laser absorbance, one can obtain $A_{FM}$ of the FM layer.

For the FM/NM bilayer:

$$Z = \frac{Z_0}{1+n+Z_0(\sigma_{FM}t_{FM}+\sigma_{NM}t_{NM})} \quad (2)$$

For the FM/NM$_1$/NM$_2$(NM$_1$/FM/NM$_2$) trilayer:

$$Z = \frac{Z_0}{1+n+Z_0(\sigma_{FM}t_{FM}+\sigma_{NM1}t_{NM1}+\sigma_{NM2}t_{NM2})} \quad (3)$$

The THz radiation impedance, Z, can be obtained by measuring the THz transmission of the sample and the bare substrate.

Normalization of the THz signals can be performed using the equation:

$$E = \frac{E_{THz}}{A_{FM}Z} \quad (4)$$

Furthermore, the relationship between the THz signal ($E_{THz}$) and the efficiency of the spin (orbit)-charge conversion can be described by adding the orbit part using the equation[9]:

$$E = \frac{E_{THz}}{A_{FM}Z} \propto (\theta_{SH}J_S + \theta_{OH}J_L) \quad (5)$$

where $\theta_{SH}$ and $\theta_{OH}$ are the spin Hall angle and the orbit Hall angle that characterize the efficiency of spin-charge and orbit-charge conversions, respectively. By using the above equation, the normalized THz signal can be shown to be proportional to the spin/orbit-charge conversion efficiency. Therefore, the spin/orbit-charge conversion efficiency of the NM layers can be estimated through the normalized THz signals.

## DATA AVAILABILITY

The data that support the findings of this study are available from the corresponding author upon reasonable request.

## ACKNOWLEDGMENTS

This work was partially supported by the National Natural Science Foundation of China (Grant Nos. 52271240, 51731003, 52061135205, 51971023, 51971024, 51927802, 52271186, 52201292) and Beijing Natural Science Foundation Key Program (Grant No. Z190007). Z.F. acknowledges the National Natural Science Foundation of China (NSFC) (62027807) and the National Key R&D Program of China (2021YFA1401400). We would like to thank the Analytical & Testing Center of Tiangong University.

## COMPETING INTERESTS

The authors declare no competing interests.

## AUTHOR CONTRIBUTIONS

P.W., Z.F., Y.H.Y., and D.L.Z. contributed equally to this work. D.L.Z initialized and conceived this work. Y.J. coordinated and supervised the project. D.L.Z. and P.W. conceived the experiments and designed all the samples. Y.H.Y. prepared the samples, Z.F. and Q.C.L.

performed the THz characterization. Z.D.X., Z.Y.J., Y.W., G.Q.Y. and X.G.X. characterized the basic properties of the samples. P.W. and D.L.Z. wrote the manuscript. All the authors discussed the results and commented on the manuscript.

**ADDITIONAL INFORMATION**

**Supplementary information** The online version contains supplementary material available.

**REFERENCES**

1. Hirsch, J. E. Spin Hall Effect. *Phys. Rev. Lett.* **83**, 1834 (1999).

2. Ando, K. et al. Inverse spin-Hall effect induced by spin pumping in metallic system. *J. Appl. Phys.* **109**, 103913 (2011).

3. Jungwirth, T., Wunderlich, J. & Olejnḱ, K. Spin Hall effect devices. *Nat. Mater.* **11**, 382-390 (2012).

4. Liu, L. Q. et al. Spin-torque switching with the giant spin Hall effect of Tantalum. *Science* **336**, 555-558 (2012).

5. Bernevig, B. A., Hughes, T. L. & Zhang, S. C., Orbitronics: the intrinsic orbital current in p-doped silicon. *Phys. Rev. Lett.* **95**, 066601 (2005).

6. Kontani, H., Tanaka, T., Hirashima, D. S., Yamada, K. & Inoue, J., Giant orbital Hall effect in transition metals: origin of large spin and anomalous Hall effects. *Phys. Rev. Lett.* **102**, 016601 (2009).

7. Go, D., Jo, D., Kim, C. & Lee, H. -W., Intrinsic spin and orbital Hall effects from orbital texture. *Phys. Rev. Lett.* **121**, 086602 (2018).

8. Jo, D., Go, D. & Lee, H. -W. Gigantic intrinsic orbital Hall effects in weakly spin-orbit coupled metals. *Phys. Rev. B* **98**, 214405 (2018).

9. Seifert, T. et al. Efficient metallic spintronic emitters of ultrabroadband terahertz radiation. *Nat. Photonics* **10**, 483-488 (2016).

10. Sinova, J., Valenzuela, S. O., Wunderlich, J., Back, C. H. & Jungwirth, T., Spin Hall effects. *Rev. Mod. Phys.* **87**, 1213 (2015).

11. Mellnik, A. R. et al. Spin-transfer torque generated by a topological insulator. *Nature* **511**, 449-451 (2014).

12. DC, M. et al. Room-temperature high spin-orbit torque due to quantum confinement in sputtered $Bi_xSe_{(1-x)}$ films. *Nat. Mater.* **17**, 800-807 (2018).

13. Zhang, W. et al. Giant facet-dependent spin-orbit torque and spin Hall conductivity in the triangular antiferromagnet $IrMn_3$. *Sci. Adv.* **2**, e1600759 (2016).

14. Zhou, J. et al. Large spin-orbit torque efficiency enhanced by magnetic structure of collinear antiferromagnet IrMn. *Sci. Adv.* **5**, eaau6696 (2019).

15. Kageyama, Y. et al. Spin-orbit torque manipulated by fine-tuning of oxygen-induced orbital hybridization. *Sci. Adv.* **5**, eaax4278 (2019).

16. Wu, Y. et al. High-performance THz emitters based on ferromagnetic/nonmagnetic heterostructures. *Adv. Mater.* **29**, 1603031 (2017).

17. Chen, X. et al. Generation and control of terahertz spin currents in topology-induced 2D ferromagnetic $Fe_3GeTe_2/Bi_2Te_3$ heterostructures. *Adv. Mater.* **34**, 2106172 (2022).

18. Feng, Z. et al. Highly efficient spintronic terahertz emitter enabled by metal-dielectric photonic crystal. *Adv. Optical. Mater.* **6**, 1800965 (2018).

19. Choi, Y. G. et al. Observation of the orbital Hall effect in a light metal Ti. https://doi.org/10.48550/ arXiv.2109.14847 (2021).

20. Lee, D. et al. Orbital torque in magnetic bilayers. *Nat. Commun.* **12**, 6710 (2021).

21. Sala, G. & Gambardella, P. Giant orbital Hall effect and orbital-to-spin conversion in 3*d*, 5*d*, and 4*f* metallic heterostructures. *Phys. Rev. Res.* **4**, 033037 (2022).

22. Lee, S. et al. Efficient conversion of orbital Hall current to spin current for spin-orbit torque switching. *Commun. Phys.* **4**, 234 (2021).

23. Go, D. & Lee, H. -W. Orbital torque: torque generation by orbital current injection. *Phys. Rev. Res.* **2**, 013177 (2020).

24. Ding, S. et al. Harnessing orbital-to-spin conversion of interfacial orbital currents for efficient spin-orbit torques. *Phys. Rev. Lett.* **125**, 177201 (2020).

25. Cong, K. et al. Coherent control of asymmetric spintronic terahertz emission from two-dimensional hybrid metal halides. *Nat. Commun.* **12**, 5744 (2021).

26. Agarwal, P. et al. Electric-field control of nonlinear THz spintronic emitters. *Nat. Commun.* **13**, 4072 (2022).

27. Martini, M. et al. Engineering the spin-orbit-torque efficiency and magnetic properties of Tb/Co ferrimagnetic multilayers by stacking order. *Phys. Rev. Appl.* **17**, 044056 (2022).

28. Hayashi, H. et al. Observation of long-range orbital transport and giant orbital torque. *Commun. Phys.* **6**, 32 (2023).

29. Seifert, T. S. et al. Time-domain observation of ballistic orbital-angular-momentum currents with giant relaxation length in tungsten. https://doi.org/10.48550/arXiv.2301.00747 (2023).

30. Du, C., Wang, H., Yang, F. & Hammel, P. C., Systematic variation of spin-orbit coupling with *d*-orbital filling: large inverse spin Hall effect in 3*d* transition metals. *Phys. Rev. B* **90**, 140407(R) (2014).

31. Xu, Y. et al. Inverse orbital hall effect discovered from light-induced terahertz emission. https://doi.org/10.48550/arXiv.2208.01866 (2022).

32. Zhang, H., et al., Laser pulse induced efficient terahertz emission from Co/Al heterostructures. *Phys. Rev. B* **102**, 024435 (2020).

33. Pai, C. F. et al. Spin transfer torque devices utilizing the giant spin Hall effect of tungsten. *Appl. Phys. Lett.* **101**, 122404 (2012).

34. Obstbaum, M. et al. Inverse spin Hall effect in $Ni_{81}Fe_{19}$/normal-metal bilayers. *Phys. Rev. B* **89**, 060407(R) (2014).

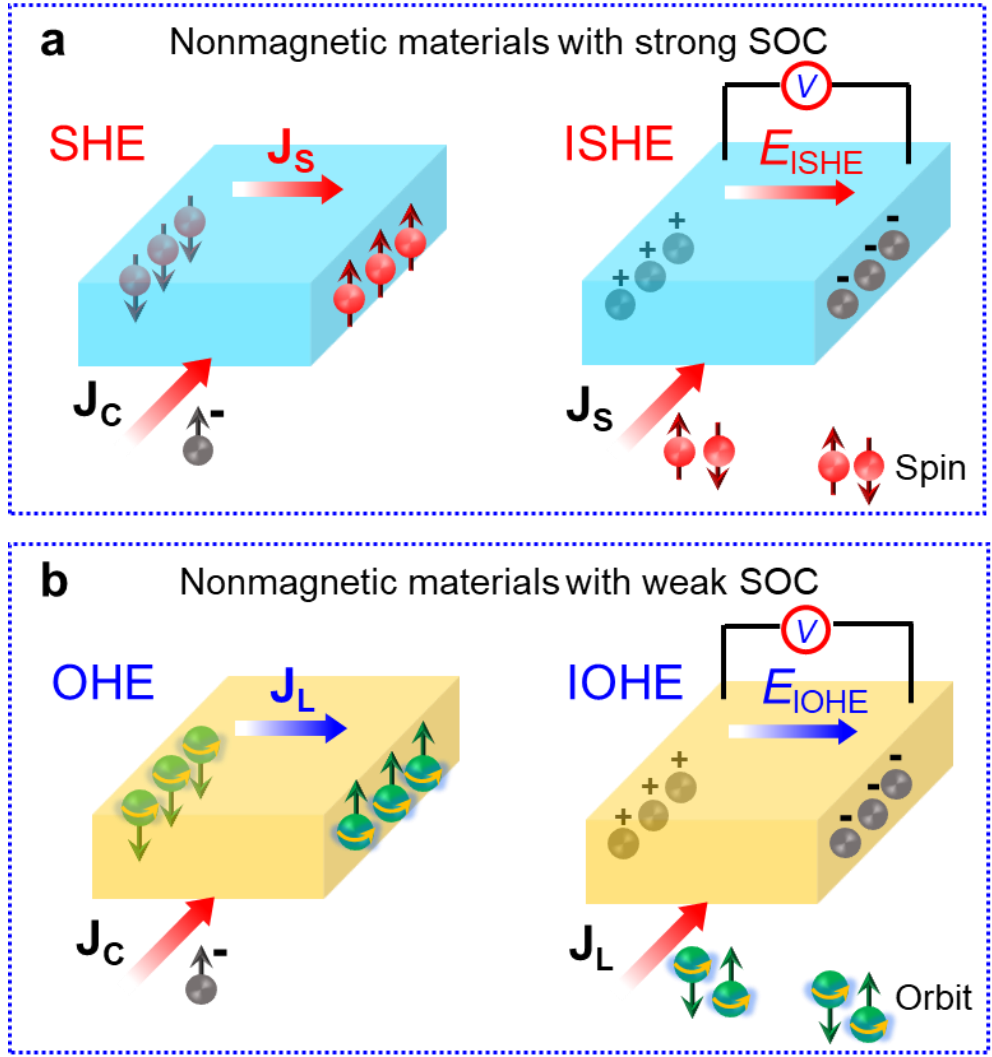

**Fig. 1** Physical mechanism of SHE/ISHE and OHE/IOHE. **a** SHE and ISHE refer to the conversions of $J_C \rightarrow J_S$ and $J_S \rightarrow J_C$ in the heavy metals or quantum materials with strong spin-orbit coupling (SOC), where a transverse flow of spin angular momentum and voltage are generated, respectively. **b** OHE and IOHE are the conversion of $J_C \rightarrow J_L$ and $J_L \rightarrow J_C$ in the materials with weak SOC, where a transverse flow of orbital angular momentum and voltage are induced, respectively.

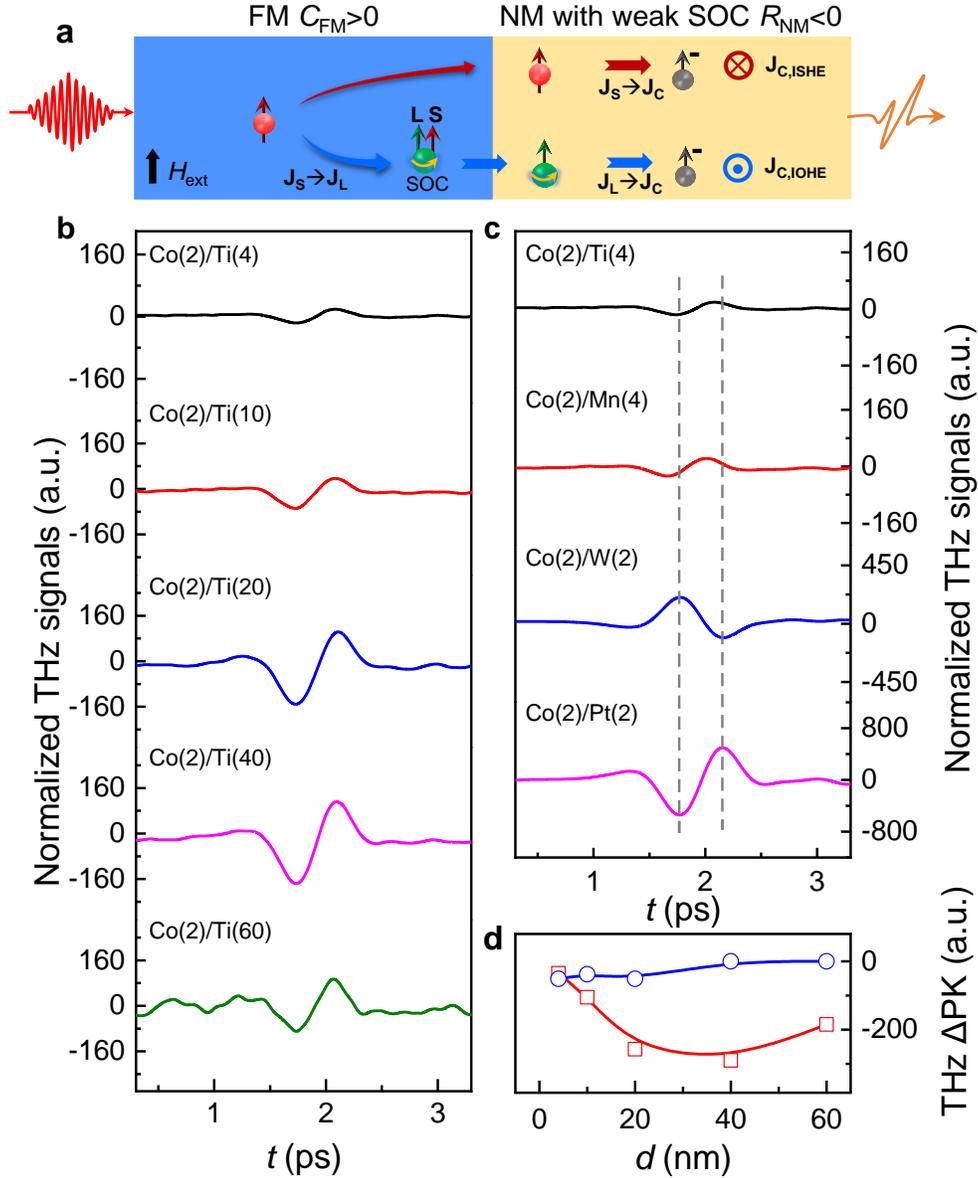

**Fig. 2** Orbitronic THz emission originated from IOHE. **a** Conversion path of $\mathbf{J_S} \rightarrow \mathbf{J_C}$ in the structure with ferromagnetic (FM) and nonmagnetic (NM) materials with weak spin-orbit coupling (SOC). When a femtosecond laser pumps into the FM layer, $\mathbf{J_S}$ and $\mathbf{J_L}$ are simultaneously generated because of SOC of the Co layer. $\mathbf{J_L}$ will flow into the NM layer and convert to $\mathbf{J_C}$ due to IOHE (labeled by the blue arrow); $\mathbf{J_S}$ will directly convert to $\mathbf{J_{C,ISHE}}$ via ISHE (labeled by the red arrow). **b** The normalized THz signals of the Co (2 nm)/Ti (4-60 nm) structures. The obvious THz signals can be observed in the Co/Ti structures, indicating the conversion of $\mathbf{J_S} \rightarrow \mathbf{J_C}$. **c** Comparison of the THz polarity of the Co/Ti, Co/Mn, Co/W and Co/Pt

structures. The Co/Ti and Co/Mn structures have the opposite and same THz polarity to those of the Co/W and Co/Pt structures, respectively. **d** The THz ΔPK of the Co/Ti (red color) and Co/Mn (blue color) structures as a function of the thickness of the Ti and Mn layers, indicating the long orbital diffusion length of Ti.

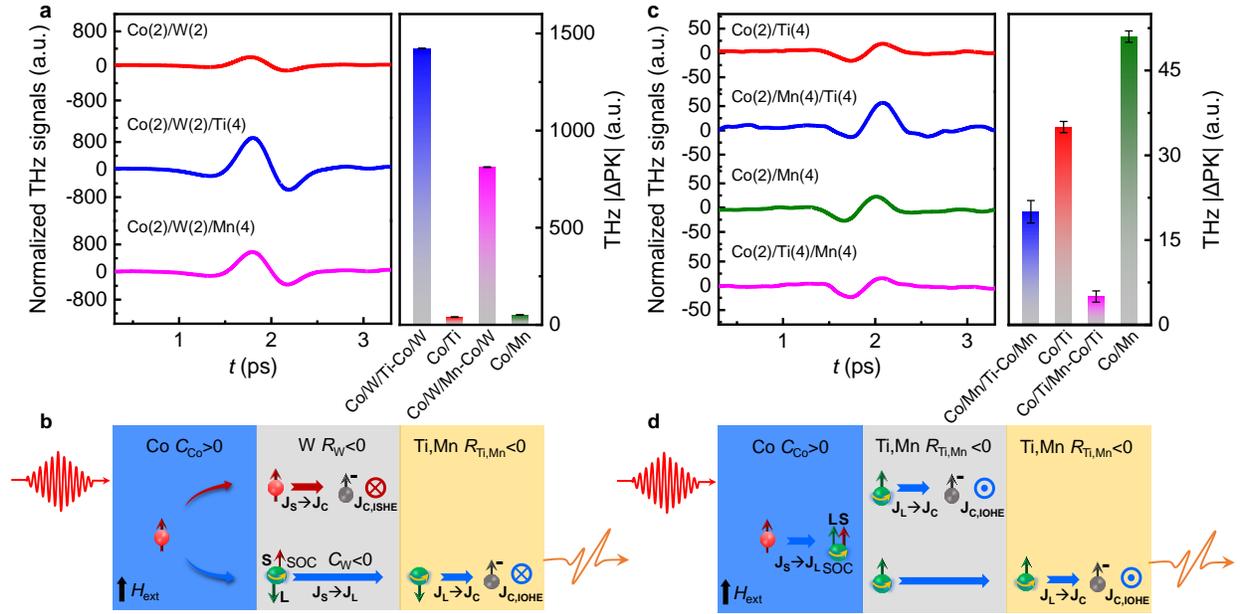

**Fig. 3** Enhanced orbitronic THz emission assisted by efficient spin-orbital current conversion. **a** The normalized THz signals of the Co/W and Co/W/$X$ ($X$ = Ti, Mn) structures and the corresponding extracted difference of THz |ΔPK|. 2-nm W inserted with strong SOC can enhance the THz amplitudes of the Co/W/$X$ ($X$ = Ti, Mn) structures compared to the Co/W and Co/$X$ ($X$ = Ti, Mn) structures. **b** The W layer provides an additional large $J_L$, then enters into the Ti or Mn layers and converts to $J_{C,IOHE}$ by IOHE. **c** The normalized THz signals of the Co/$X$ ($X$ = Ti, Mn), Co/Ti/Mn, and Co/Mn/Ti structures and the corresponding extracted difference of THz |ΔPK|. The difference of THz |ΔPK| of the Co/Ti/Mn, Co/Mn/Ti, and Co/$X$ ($X$ = Ti, Mn) structures are smaller than those of the Co/W/$X$ ($X$ = Ti, Mn) structures. **d** The routine conversion of $J_S \rightarrow J_L$ by SOC of the Co layer and $J_L \rightarrow J_C$ by IOHE of the Ti and Mn layers. The error bars refer to the noise of the THz signals.

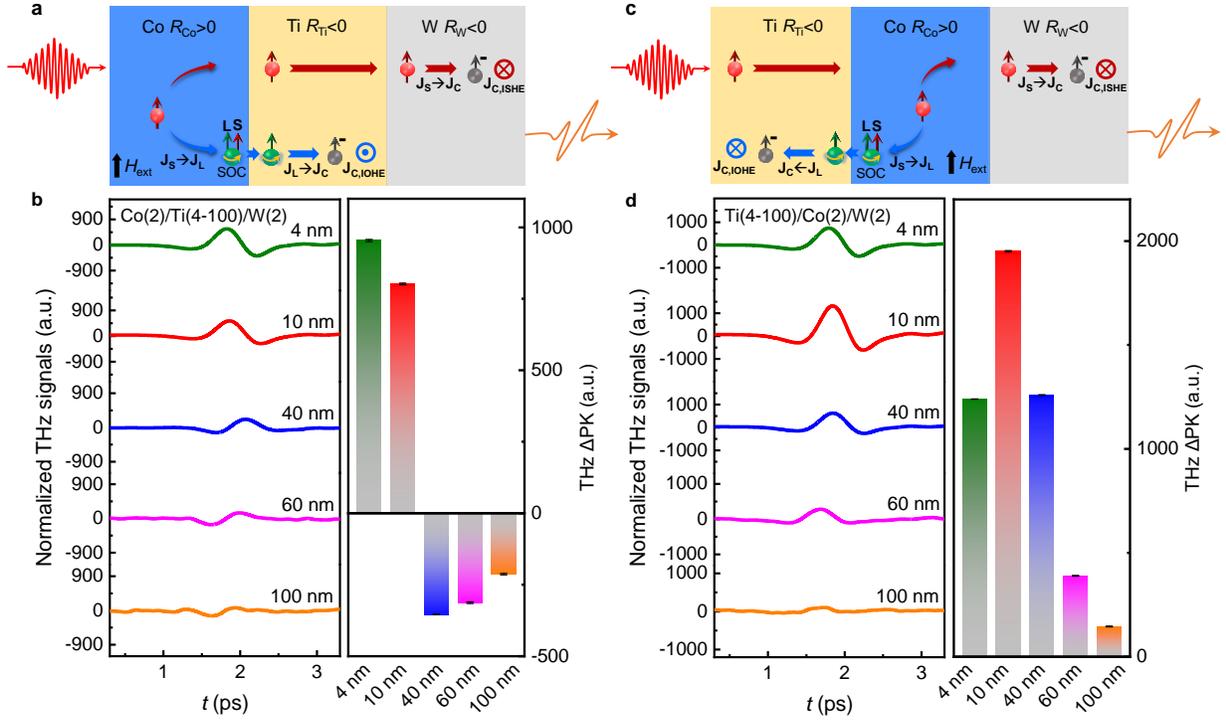

**Fig. 4** Collaboration between IOHE and ISHE. **a** The $J_S{\rightarrow}J_L{\rightarrow}J_C$ and $J_S{\rightarrow}J_C$ conversion of the Co/Ti/W structure. $J_S$ from the Co layer will be separated into two parts, one part will transfer into the W layer through the Ti layer and convert to $J_{C,ISHE}$ due to ISHE; the other part will convert to $J_L$ and flow into the Ti layer, then convert to $J_{C,IOHE}$ due to IOHE. $J_L$ from the W layer is not induced. **b** The normalized THz signals of the Co (2 nm)/Ti (4-100 nm)/W (2 nm) structures and the corresponding extracted THz ΔPK. **c** The $J_S{\rightarrow}J_L{\rightarrow}J_C$ and $J_S{\rightarrow}J_C$ conversion of the Ti/Co/W structure. The THz signals originate from $J_S{\rightarrow}J_{C,W}$ by ISHE of the W layer and $J_S{\rightarrow}J_L{\rightarrow}J_{C,Ti}$ by IOHE of the Ti layer. **d** The normalized THz signals of the Ti (4-100 nm)/Co (2 nm)/W (2 nm) structures and the corresponding extracted THz ΔPK. The error bars refer to the noise of the THz signals.

19